\documentstyle[preprint,aps]{revtex}
\begin{document}
\draft
\title {Complex ion formation in liquid Ag-Se alloys}
\author{F. Kirchhoff, J.  M.  Holender and M.  J.  Gillan}
\address{ Physics Department, Keele University \\
Keele, Staffordshire ST5 5BG, U.K.}
\date{\today}
\maketitle
\begin{abstract}
Ab initio molecular dynamics simulations are used to investigate the
structure and electronic properties of the liquid Ag-Se system at
three compositions. The realism of the simulations is demonstrated by
comparison with diffraction data for the stoichiometric case Ag$_2$Se. As
the Se content is increased beyond the stoichiometric value,
short-lived Se$_n$ complexes are formed. The concentration of complexes
and the associated changes of electronic structure can
be explained using a simple ionic model.
\end{abstract}
\narrowtext

Many binary liquids show a dramatic dependence of their
structure and electronic properties on composition \cite{enderby90}.
A celebrated
example is the Cs-Au system, in which the pure elements are
excellent metals, but nevertheless the equiatomic mixture has a very
low conductivity and a structure characteristic of a molten
salt\cite{hensel79}.
These effects arise from the electronegativity difference between the
elements, and the resulting charge transfer, partial ionicity
and atomic ordering.
In systems where one of the elements is a semiconductor in the liquid
state -- for example alloys of metals with S or Se --
even richer behavior can be expected, since variation of composition
should change the bonding from metallic through partially ionic to
covalent. We report here a set of simulations of the Ag-Se system
performed using {\em ab initio} molecular dynamics (AIMD), which
we have used to explore these effects.

The Ag-Se system has been
studied by a variety of experimental techniques \cite{endo80},
but understanding
of its structure is limited,
since diffraction measurements have been made only
near the stoichiometric composition Ag$_2$Se\cite{price93}, and these
yield only the {\em total} structure factor and not the partial correlations
between the two elements.
The present simulations allow us to build up a detailed picture of
the composition-dependent structure, and also
give important new insights into the electronic
structure and into dynamical aspects of the bonding, which would be hard to
probe experimentally. A number of previous AIMD simulations of liquid
metals and semiconductors have been reported\cite{sti89},
but the Ag-Se system presents a significant technical challenge,
as will become clear.

We shall show that our simulations are in excellent agreement with
available structural
measurements at the stoichiometric composition Ag$_2$Se, where we find
that a partially ionic description is appropriate. As soon as the Se
content is increased beyond the stoichiometric value, a major
structural change begins, consisting of formation of Se$_n$ complexes.
This is accompanied by changes of the electronic density of states
(DOS) associated with the formation of covalent Se-Se bonds. Both the
concentration of Se$_n$ complexes and the changes in the DOS can be
understood using a simple ionic model. In spite of their major
structural role, the complexes turns out to be surprisingly
short-lived, the typical bond lifetime being $\sim$ 0.5~ps.

Our AIMD technique, like the methods pioneered by Car and
Parrinello \cite{cp85}, uses density functional theory, pseudopotentials,
and a plane-wave representation of the orbitals \cite{df}. However, instead of
treating the electronic degrees of freedom as fictitious dynamical
variables, we use conjugate-gradient minimization \cite{pay92} to reach the
self-consistent ground state at each step. The
Hellmann-Feynman forces are then used to integrate the classical equation
of motion of the ions.  To handle the semi-metallic
nature of the system, we use Fermi-surface smearing, with
the electronic occupation numbers treated as auxiliary dynamical
variables \cite{gil89,kre94,gru94}.

The details of our calculations are as follows.
We use {\em ab-initio} norm-conserving non-local pseudopotentials, with
exchange and correlation included {\em via} the local density
approximation in Ceperley-Alder form\cite{lda}.
The Ag pseudopotential has been optimized using the method of
Lin {\em et al.} \cite{lin}, which is a refinement of the scheme due to
Rappe {\em et al.} \cite{rap}.  The Se pseudopotential does not require
optimization and the standard Kerker method \cite{ker} suffices.
The $s$ and $p$ components of the Ag
pseudopotential were generated using the atomic configuration
4$d^{10}$ 5$s^{0.25}$ 5$p^{0.25}$, and the $d$ component using the
configuration 4$d^{10}$ 5$s^{0.5}$. The core radii were 2.0, 2.0 and
2.5 a.u. for the $s$, $p$ and $d$ components respectively.
For Se we used 4$s^{2}$ 4$p^{4}$ for the $s$ and $p$ waves
and 4$s^{2}$ 4$p^{2.75}$ 3$d^{0.25}$ for the $d$ wave; the core radii for
were chosen to be 2.0, 2.0, and 2.3 a.u. for the $s$, $p$ and $d$
components  respectively.
We use the pseudopotentials in Kleinman-Bylander separable form \cite{kb}
with the $s$-wave treated as local;
the non-local parts of the pseudopotentials are treated in real
space \cite{kin91}.
Tests on the low-temperature crystal phase of Ag$_2$Se show that these
pseudopotentials reproduce the experimental structure very well and
that a plane-wave cutoff of 400~eV is sufficient to achieve good
convergence of the total energy; this cutoff is used throughout the
present work.
We have also tested these potentials on AgCl \cite{agcl}, pure Se and
GaSe. In every case we find very good agreement with
the experimental structure.

AIMD calculations on systems containing post-transition metals such
as Ag are extremely demanding, because it is essential to include
the $d$-electrons as valence electrons, and because of the large
plane-wave basis set needed to represent the $d$-orbitals.
As a result, very few such AIMD simulations have been
reported \cite{pas92,kre93}.
The present work relies heavily on parallel computation methods,
and has been performed using the parallel {\em ab initio} code
CETEP \cite{cetep} running on a Cray T3D machine.

Our simulations of the Ag-Se liquid alloys have been performed
on a system of 69 atoms in a cubic box with the usual periodic
boundary conditions. We use $\Gamma$-point sampling, and
a Fermi-smearing energy width of 0.2 eV.
The Verlet algorithm is used to integrate the ionic equation of
motion, with a time step of 3~fs.
We have performed simulations at the temperature $T \simeq 1350 \; K$
for three concentrations of Ag$_{1-x}$Se$_{x}$,
namely: $x=0.33$ (46 Ag atoms and 23 Se atoms), $x=0.42$ (40 Ag atoms and 29
Se atoms) and $x=0.65$ (24 Ag atoms and 45 Se atoms).
The simulations are performed at a density which is linearly interpolated
between experimental values for $\ell$-Ag$_2$Se and $\ell$-Se.

To initiate the simulations, we exploit the fact that
an empirical pair-potential model \cite{rino88} has been developed for the
stoichiometric Ag$_2$Se system, which gives a reasonable
reproduction of the liquid structure when used in classical
molecular dynamics. We begin by making simulations with this
empirical model, and we then switch over to AIMD and let the system
equilibrate for a further 1~ps before collecting data over the
next 2~ps. We reach the other Ag$_{1-x}$Se$_x$ compositions by
replacing some of the Ag atoms by Se atoms and then equilibrating
for 1~ps at the new composition; production runs of 2~ps are
again performed in each case.

Fig.\ \ref{nsf} shows the neutron-weighted
static structure factor $S(k)$
obtained from our simulation for the Ag$_2$Se composition,
compared with experimental data at $T$~=~1150~K \cite{enderby90}.
The overall agreement between theory and experiment is excellent, with
all the main features having the correct position and
magnitude.  It is also
remarkable that our simulation reproduces the position of
the small pre-peak at wave vector $k \simeq 1.7$~\AA$^{-1}$;
its height, however,
is lower than the experimental value. This discrepancy is probably
due to the finite size of our simulation cell and the resulting
lack of $k$-space resolution. Our simulation allows us to analyze
the origin of the peaks in $S(k)$ in terms of the
calculated partial structure factors $S_{\mbox{\scriptsize Ag-Ag}}(k)$,
$S_{\mbox{\scriptsize Ag-Se}}(k)$ and $S_{\mbox{\scriptsize
Se-Se}}(k)$. We find that
the main peak in $S(k)$ at $k \simeq 2.7$~\AA$^{-1}$ is
due to peaks
in $S_{\mbox{\scriptsize Ag-Ag}}(k)$ and
$S_{\mbox{\scriptsize Ag-Se}} (k)$, which reinforce each other.
The pre-peak in $S(k)$ arises from a positive peak in
$S_{\mbox{\scriptsize Se-Se}}(k)$ and a negative peak in
$S_{\mbox{\scriptsize Ag-Se}}(k)$, which partially cancel each other.
A more detailed analysis of the structure factors will be
given elsewhere.

The partial pair correlation functions $g_{\mbox{\scriptsize Ag-Ag}}(r)$,
$g_{\mbox{\scriptsize Ag-Se}}(r)$ and $g_{\mbox{\scriptsize Se-Se}}(r)$ for
the three concentrations are displayed in Fig.\ \ref{rdf}.
The results show that increase of Se content causes dramatic changes
in $g_{\mbox{\scriptsize Se-Se}}(r)$.
As Se content increases,
a short-distance peak builds up.
At stoichiometry, $g_{\mbox{\scriptsize Se-Se}}(r)$ exhibits broad
peaks at 3.99 and 4.83\AA, and only a very weak tail below
3.0~\AA. For $x=0.42$, instead of a tail, there is a
short-distance peak at 2.35~\AA. The position of the main peak has shifted
to 4.72~\AA\ and has decreased in magnitude. The peak around 4.0~\AA\ seen
at stoichiometry has merged with the main peak to give rise
to a shoulder. At the last concentration, the short-distance peak
is dominant and $g_{\mbox{\scriptsize Se-Se}}(r)$ shows little
structure beyond 4.0~\AA\, with a low and broad second peak.
We note that the radius 2.35~\AA\ associated with the short-distance
peak is very close to the Se-Se covalent bond length in
crystalline and liquid Se\cite{suzuki76}. The growth of the short-distance peak
can
be characterized by the Se-Se coordination number within a sphere
of radius 3.0~\AA, which we calculate to be 0.1, 0.71 and 1.7 for
$x$ = 0.33, 0.42 and 0.65 respectively.
By contrast with the major changes in $g_{\rm Se-Se} (r)$,
the other two
correlation functions change rather little
with composition.

More insight into the structural changes can be gained from
`snapshots' of the ionic positions at the three concentrations
(Fig.\ \ref{snap}). To aid the eye, bonds have been drawn between Se
atoms separated by less than 3.0~\AA. As expected from the
results for $g_{\alpha \beta} (r)$, the major feature is the
formation of Se clusters as $x$ exceeds 0.33. Already at
$x = 0.42$, the Se atoms bond not only into dimers but also into
larger Se$_n$ clusters. At this composition, only $\sim$~48~\%
of the Se atoms are bonded, 76~\% of the bonded atoms being
one-fold coordinated and 22~\% two-fold, the remaining 2~\% having
higher coordination. The dominance of one-fold and two-fold
coordination means that most of the clusters are either dimers or
Se$_n$ {\em chains}.

At the composition $x = 0.65$, most of the Se atoms are in clusters,
with only 7~\% being isolated. The proportions of bonded Se in
one-fold and two-fold coordination are now 35~\% and 40~\%,
with a non-negligible 15~\% being three-fold coordinated. The
liquid is thus composed of one-dimensional Se$_n$ chains
interconnected {\em via} three-fold coordinated atoms, as shown in
Fig.\ \ref{snap}c. The Se$_n$ clusters at this composition are large,
with $n$ typically $> 20$.

The structural changes are intimately linked to the electronic
structure. The calculated electronic density of states (DOS) and the local
DOS on Ag and Se atoms for the three compositions are shown in Fig. 4.
The main features are Se(4$s$) states at --12~eV, Ag(4$d$) states at --4~eV
and Se(4$p$) states in the region above --7~eV; Ag(5$s$-$p$) states extend
upwards from roughly the Fermi level $E_{\rm F}$, and there is a significant
hybridization between these and the Se(4$p$) states. The Se(4$p$) band
of states stands out clearly from other parts of the DOS at all
three compositions. For Ag$_2$Se, $E_{\rm F}$ lies in a pseudo-gap at the
top of this band; the number of occupied 4$p$ states per Se atom is equal
to 3, and a partially ionic model approximating to Ag$_2^+$Se$^=$ is
appropriate. As $x$ increases beyond 0.33, the number of occupied 4$p$
states per Se atom falls below 3, and we attribute this to the formation
of unoccupied anti-bonding 4$p$ combinations associated with clusters.
These anti-bonding states are hybridized with Ag(5$s$-$p$) states,
and only become clearly visible at $x = 0.65$.
We also note that with increasing $x$ the Se(4$s$) band broadens strongly,
and we can show that this arises from the formation of bonding and
anti-bonding Se(4$s$) combinations in the clusters.

Our analysis of the structure and the DOS thus leads to a simple
approximate picture of liquid Ag-Se alloys. At the Ag$_2$Se
composition, the system consists approximately of isolated
Ag$^+$ and Se$^=$ ions. Our picture is that
as the Se content increases, chain-like
Se$_n$ clusters are formed, and the system consists of
a mixture of Ag$^+$, Se$^=$ ions and (Se$_n$)$^=$ complexes.
As confirmation of this picture, we can use charge-balance arguments
to estimate the number of Se-Se bonds at any composition.
If the system consisted entirely of Ag$^+$ and Se$^=$ ions, the
net charge on a system of $N_{\rm Ag}$ and $N_{\rm Se}$ ions
would be $N_{\rm Ag} - 2 N_{\rm Se}$. If the clusters are all
(Se$_n$)$^=$, the formation of every Se-Se
bond reduces the charge by two units. For electroneutrality,
the number of bonds must therefore be $N_{\rm Se} - \frac{1}{2}
N_{\rm Ag}$ and the coordination number of the short-distance
peak in $g_{\rm Se-Se} (r)$ must be $2 - N_{\rm Ag}/N_{\rm Se} =
3 - 1/x$. This gives predicted values of 0, 0.62 and 1.46 for the
selenium-selenium
coordination number at the three compositions, which are quite close to
the values 0.1, 0.71 and 1.7 reported above. There are experimental
indications for the occurrence of complex ions in other
chalcogenide-metal alloys, for example
(Se$_2$)$^=$ pairs in CuSe\cite{cuse}
and (Te$_n$)$^=$ chains in the Te rich side of K-Te \cite{kte}.

In view of the dominant structural r\^{o}le played by Se$_n$ clusters,
the dynamics of their formation and dissolution is of great interest.
We have performed a statistical analysis on the lifetime
of Se-Se bonds, which shows that under the conditions we have studied they
have a very short half-life on the order of 0.5~ps. A striking effect
that seems to be linked with this is that the Se diffusion coefficient
increases with increasing Se content. Details of these dynamical effects
will be described elsewhere.

In conclusion, we have performed AIMD simulations on the liquid
Ag-Se system to investigate how the atomic ordering and electronic
structure evolve with composition in this typical alloy between
metallic and semiconducting elements. The close agreement with
neutron-diffraction data confirms the realism of the simulations.
We have shown that the formation of chain-like Se$_n$ complexes
begins as soon as the Se content exceeds the stoichiometric value,
and that the covalent Se-Se bond-length is very close to its
value in crystalline and liquid Se. The concentration of Se-Se
bonds and the changes of electronic structure can be understood using
a simple ionic model in which the liquid is a mixture of Ag$^+$,
Se$^=$ and (Se$_n$)$^=$ ions. Although the Se$_n$ complexes
are structurally so important, individual Se-Se bonds are extremely
transient, with a lifetime of less than a picosecond.

\acknowledgments

The work was done within the U.K. Car-Parrinello
Consortium, which is supported by the High Performance Computing
Initiative. A time allocation on the Cray T3D at EPCC is acknowledged.
Discussions with J. Enderby and A. Barnes, and
technical help from I. Bush, M. Payne, A. Simpson and J. White
played a key r\^{o}le.
JMH's work is supported by EPSRC grant GR/H67935.
The work used distributed hardware
provided by EPSRC grants GR/H31783 and GR/J36266.

\begin{figure}
\caption{The total neutron weighted structure
$S(k)$ factor of Ag$_2$Se.
Solid line and circles represent simulation and experimental results
\protect \cite{enderby90} respectively.}
\label{nsf}
\end{figure}

\begin{figure}
\caption{Partial radial distribution functions $g_{\alpha\beta}(r)$ of
$\ell$-Ag$_{1-x}$Se$_x$ at concentrations $x$=0.33 (full line), $x$=0.42
(dotted line) and $x$=0.65 (dot-dashed line).}
\label{rdf}
\end{figure}

\begin{figure}
\caption{Snapshots of typical configurations of $\ell$-Ag$_{1-x}$Se$_x$ at
concentrations (a) $x$=0.33, (b) $x$=0.42 and (c) $x$=0.65. Silver atoms
are shown as black spheres, selenium atoms as gray spheres. Bonds are drawn
between Se atoms with separation $<$ 3.0 \AA. Bonds to atoms in neighboring
cells are represented with two-colored sticks.}
\label{snap}
\end{figure}

\begin{figure}
\caption{Density of states (solid curve) and local densities of
states (LDOS) for Ag (chain curve) and Se (dotted curve) from
simulations of Ag$_{1-x}$Se$_x$ at $x$ = 0.33, 0.42 and 0.65.
For clarity, the scale used for the Se LDOS is four times that
used for the Ag LDOS. The vertical dotted line marks the Fermi
energy.}
\label{dos}
\end{figure}

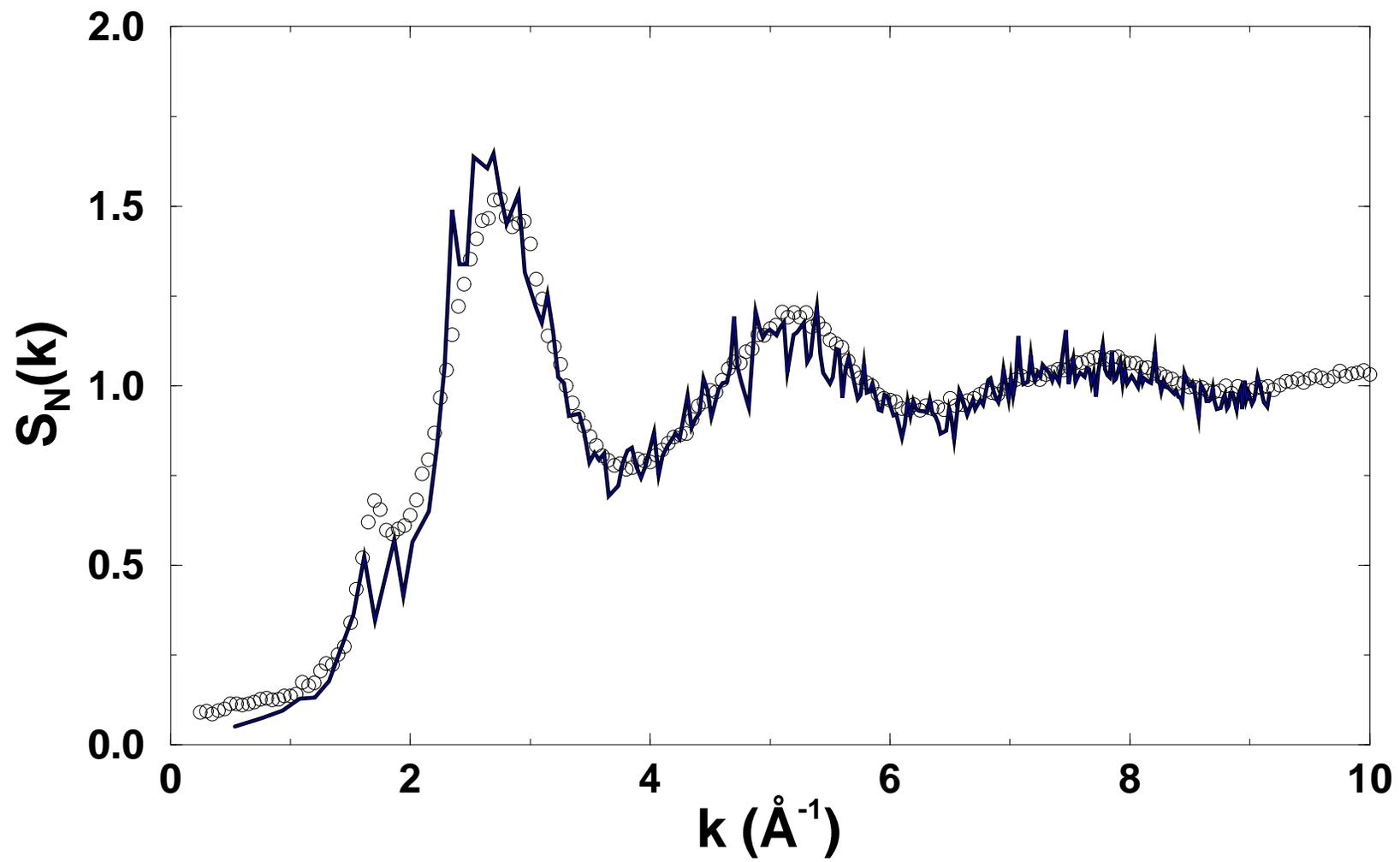
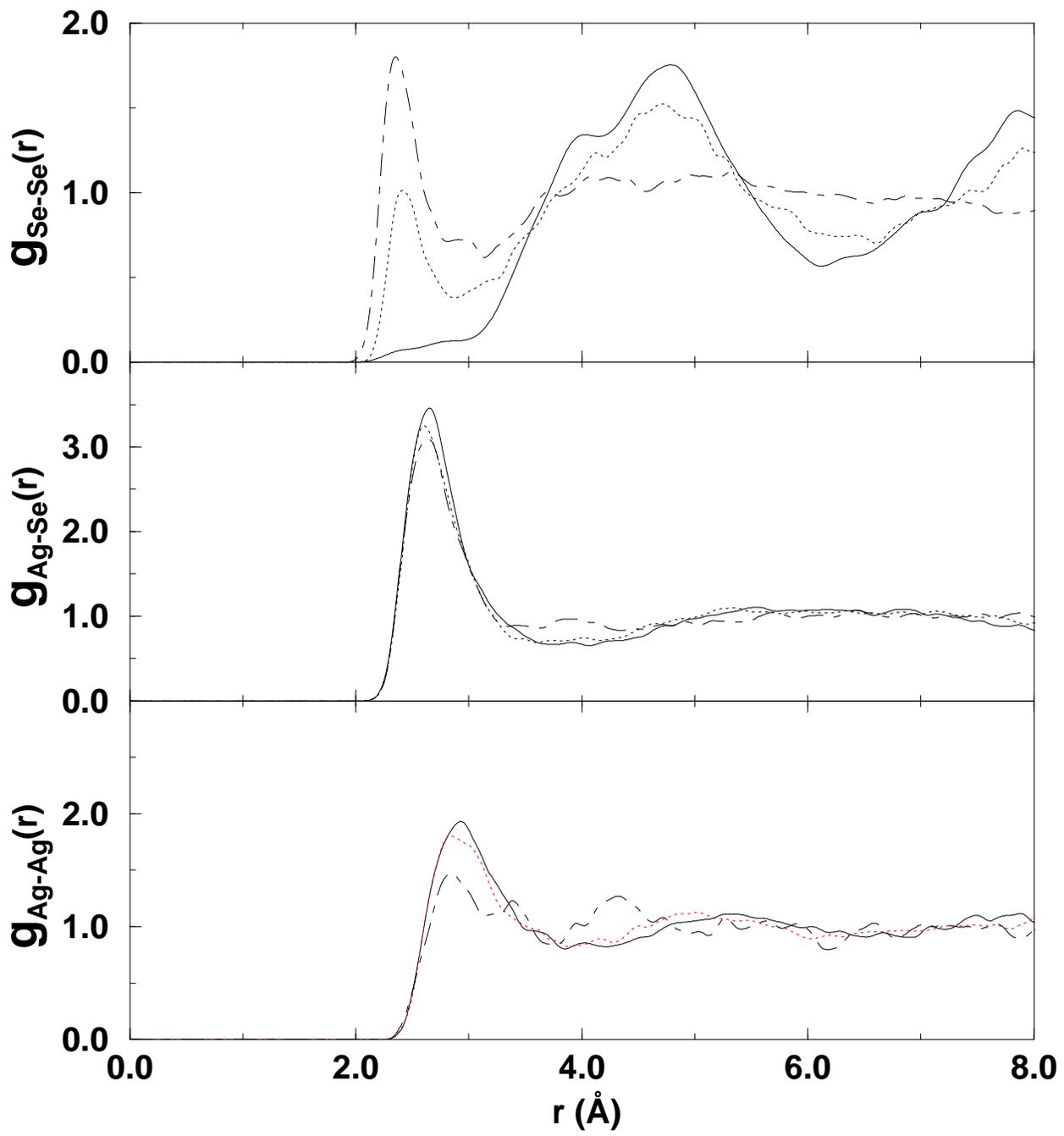
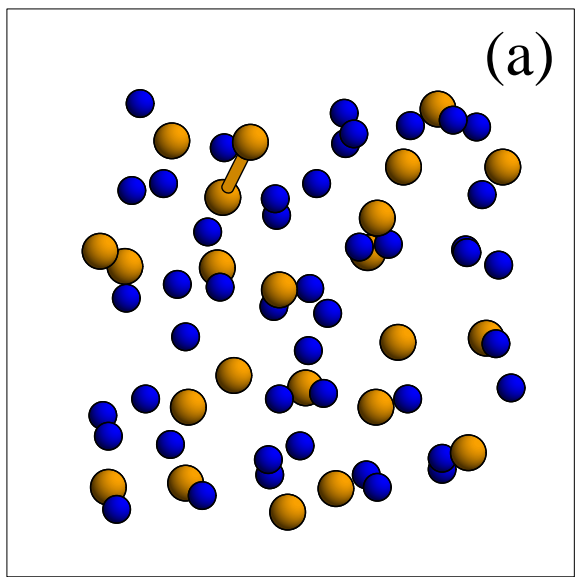
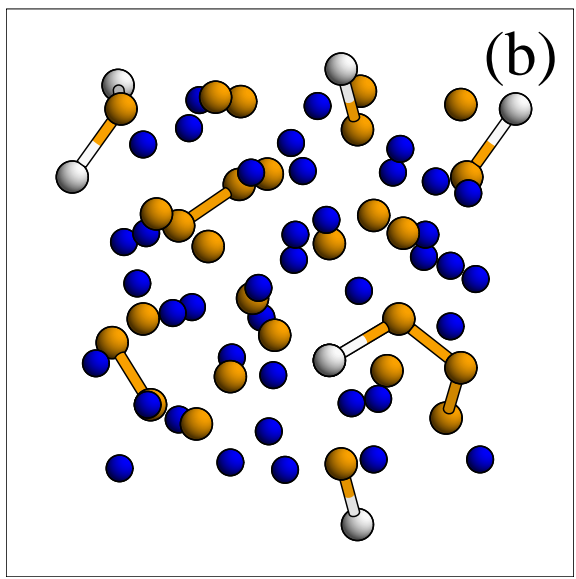
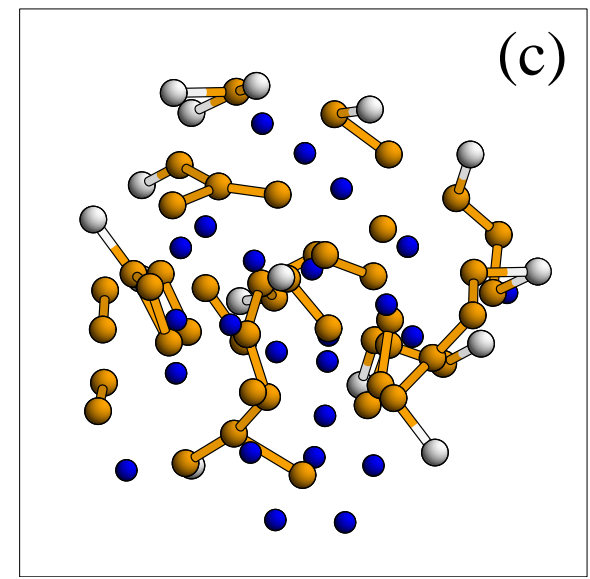
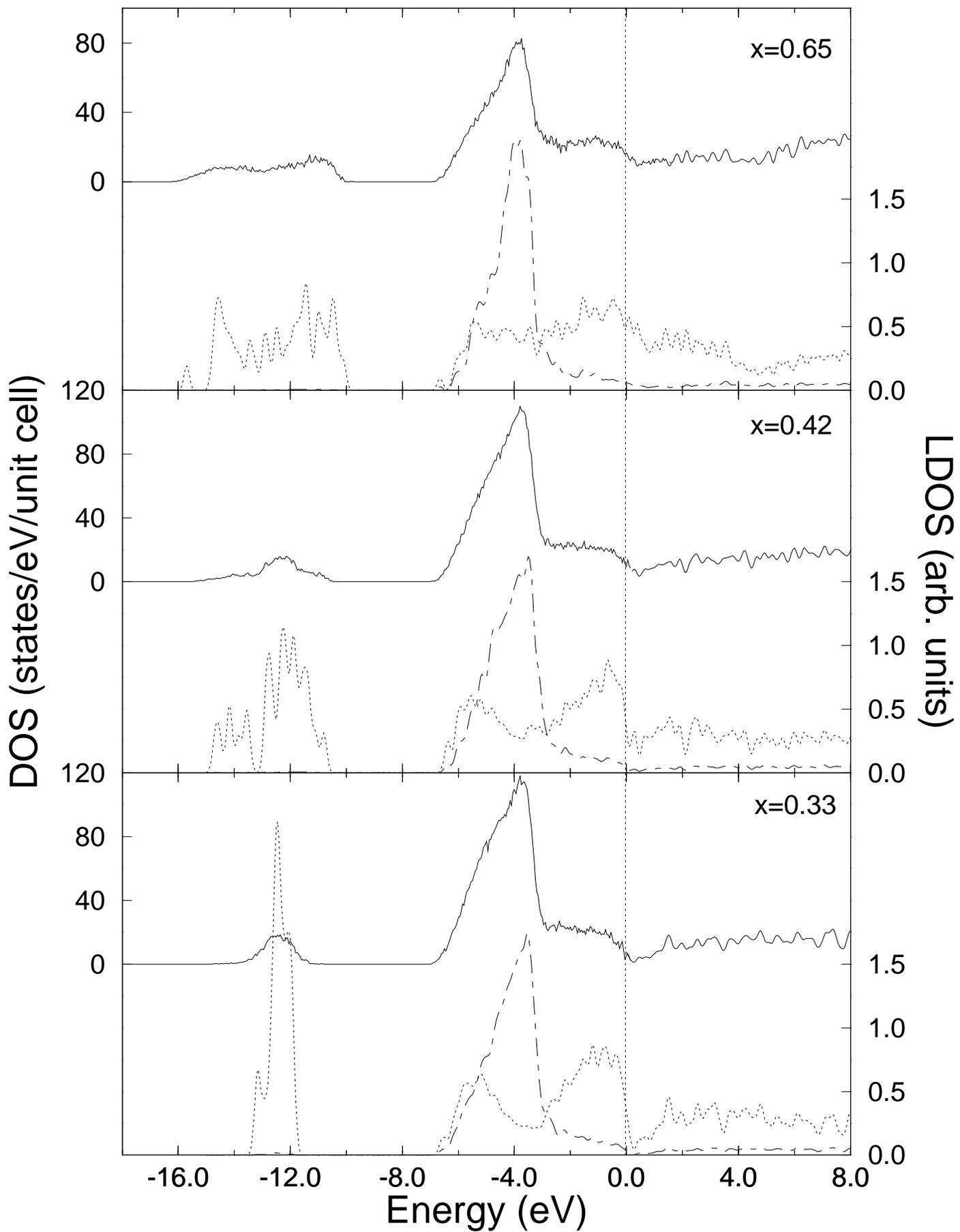
\end{document}